\documentclass[aps,twocolumn,superscriptaddress,showpacs]{revtex4}
\usepackage{bm}
\usepackage{graphicx}

\begin{document}
\bibliographystyle{revtex}


\title{Dielectric constant of glasses: first observation of a two-dimensional behavior.}

\author{F. Ladieu,$^{1,*}$ J. Le Cochec,$^1$ P. Pari,$^2$ P. Trouslard$^3$, and P. Ailloud}
\email[]{ladieu@drecam.cea.fr}
\affiliation{DSM/DRECAM/LPS, $^2$SPEC, $^3$INSTN/LVDG, C.E.Saclay, 91191 Gif/Yvette, France}

\date{\today}

\begin{abstract}
The $1$kHz real part $\chi'$ of the dielectric constant of a structural glass was measured at 
low 
temperature $T$ down to $14$ mK. Reducing the sample thickness $h$ to $10$ nm suppresses the 
usual 
minimum of $\chi'$ for measuring fields $E<.5 $ MV/m. This contradicts the Two Level System 
(TLS) 
model but is well accounted for by including TLS-TLS interactions where excitations delocalize 
between 
TLS's through a $E$-induced  mechanism recently designed: for small $h$'s this interaction is 
reduced, which  
explains the two-dimensional behavior of $\chi'(T)$. Hence, interactions play a key role in standard thick samples.

\end{abstract}
\pacs{61.43.Fs, 77.22.Gm, 72.20.Ht, 72.20.My}

\maketitle

Since the 1970's amorphous solids have been widely investigated at low temperature $T$ 
\cite{Philipps}, and their properties turned out to be both "universal" (i.e. quasi independent from 
the 
chemical composition) and strongly different from their crystalline counterparts. This 
is 
explained within the tunneling two-level system (TLS) model \cite{Anderson} where some atomic 
species  
fluctuate between two neighboring energy minima separated by a potential barrier, 
which, at low $T$ is crossed by Tunneling.

 Even if the low density of glasses (in comparison with   
their crystalline counterpart) may justify \cite{Lock} the existence of "voids" and thus the 
TLS framework, the 
ability of the standard TLS model, where the TLS's do \textit{not} mutually interact, to 
account 
for 
experiments seems very intriguing. Indeed, drawing from measurements the coupling strength of 
a TLS to 
phonons, one finds \cite{Joffrin} that TLS's are \textit{strongly coupled to each other} via 
\textit{virtual phonons}: 
a 
tunnel transition on a given TLS deforms elastically the neighboring matrix, yielding an 
energy 
change of $U \propto 1/r^3$ for a TLS located at a distance $r$, and for two neighboring 
TLS's \cite{Neu} one 
gets  $U(r=1$ nm$) \simeq 10$ K. Similarly, since many TLS's are charged, they  
interact through dipolar interaction, mediated by \textit{virtual photons}, yielding an 
interaction 
energy \cite{Neu} in the same range of $10$ K between neighboring TLS's. This large energy 
scale contrasts with 
the weakness of the rare experimental evidences of TLS's interactions: in the $100$ mK 
range, 
only small instationarities \cite{Carruzzo}, and unexpected small $H$ effects 
\cite{Strehlow}, on the kHz dielectric 
susceptibility $\chi$ have been related to interactions. In the few mK range, 
somewhat larger 
effects, such as the ultra-low-$T$ plateau \cite{Enss} in the dielectric constant, and the internal friction 
behavior \cite{Thompson}, 
were explained with 
interactions. 

This work yields \textit{strong} evidence of the key role of TLS's interactions in the $100$ mK range. Indeed, we 
show that reducing 
the thickness $h$ of glassy samples down to 
$10$ nm strongly affects the real part $\chi'$ of the $1$ kHz 
dielectric 
constant: the 
$\chi'(T)$ minimum, which occurs at a given $T_{rev}$ for $h > 100$ nm, 
is progressively \textit{moved to lower} $T$ as $h$ \textit{is decreased}, and finally disappears for  
$h 
\simeq 10$ nm. This new $\chi'(T)$ behavior contradicts the non interacting TLS model and evidences the role of 
TLS's interactions, 
since the latter are strongly reduced at small $h$'s, as we shall see. 
Within the spectrum of theories 
dealing with interactions in the TLS model, 
ranging 
from the one stating that interactions are renormalized to zero by frustration 
\cite{Coppersmith}, to the one assuming 
that interactions supersede disorder \cite{Wurger}, the mechanism 
proposed by 
Burin \textit{et al.} \cite{Burin} will prove to account for the reported data, once included in 
numerical simulations 
of $\chi'$ \cite{LecochecNL}. 

The samples were all produced \textit{similarly}: on a vitreous $a$-SiO$_2$ 
$0.1$ mm 
thick substrate, a Cu/glass/Cu/glass structure was deposited, by using a multi chamber system 
excluding exposure to air during the \textit{whole} process \cite{LecochecH}. The $15$ nm thick 
Cu electrodes were  
evaporated ($.1$nm/s), while the glass layers were made from TetraEthylOrthoSilane with a 
$13$ MHz vapor 
plasma where the autopolarisation was set at $-100$ V, the incoming flux at $2$ sccm, and the 
pressure at 
$0.80$ Pa. The resulting glass deposition rate was $.1$ nm/s, allowing to set the thickness $h$ 
of the 
internal glass layer by choosing the deposition time. Since the $h$ 
value is crucial, it was further measured by three other methods which all gave 
compatible 
results: \textit{ i)} in situ laser interferometry was realized onto a Si substrate placed 
close to the 
sample during the glass deposition; \textit{ ii)} the glass layer (the one grown onto the Si 
substrate) was 
irradiated by a deuton $.91$ MeV beam allowing, through the nuclear reaction on $_{16}$O, an 
$h$ 
estimate; \textit{ iii)} the value of the capacitance $C\propto \chi'$ was checked to scale 
with the 
expected $h$. The $15$ nm thick top glass layer hinders any spurious atmospheric effect 
during the 
cryogenic experiment. The sample was glued inside a copper box connected 
to the 
mixing chamber of the $_3$He/$_4$He dilution refrigerator. Semi-rigid coaxial shielding was 
ensured from the cold 
copper box up to the 
$2500$-Andeen capacitance bridge. A capacitance $c_F$, twice larger than that of the sample, 
was set 
in parallel of each cable, so as to filter  
high frequency 
parasitic fields: the data were unchanged when $c_F$ was halved, proving the filtering efficiency.

\begin{figure}
\includegraphics[height=6cm, width=8.5cm]{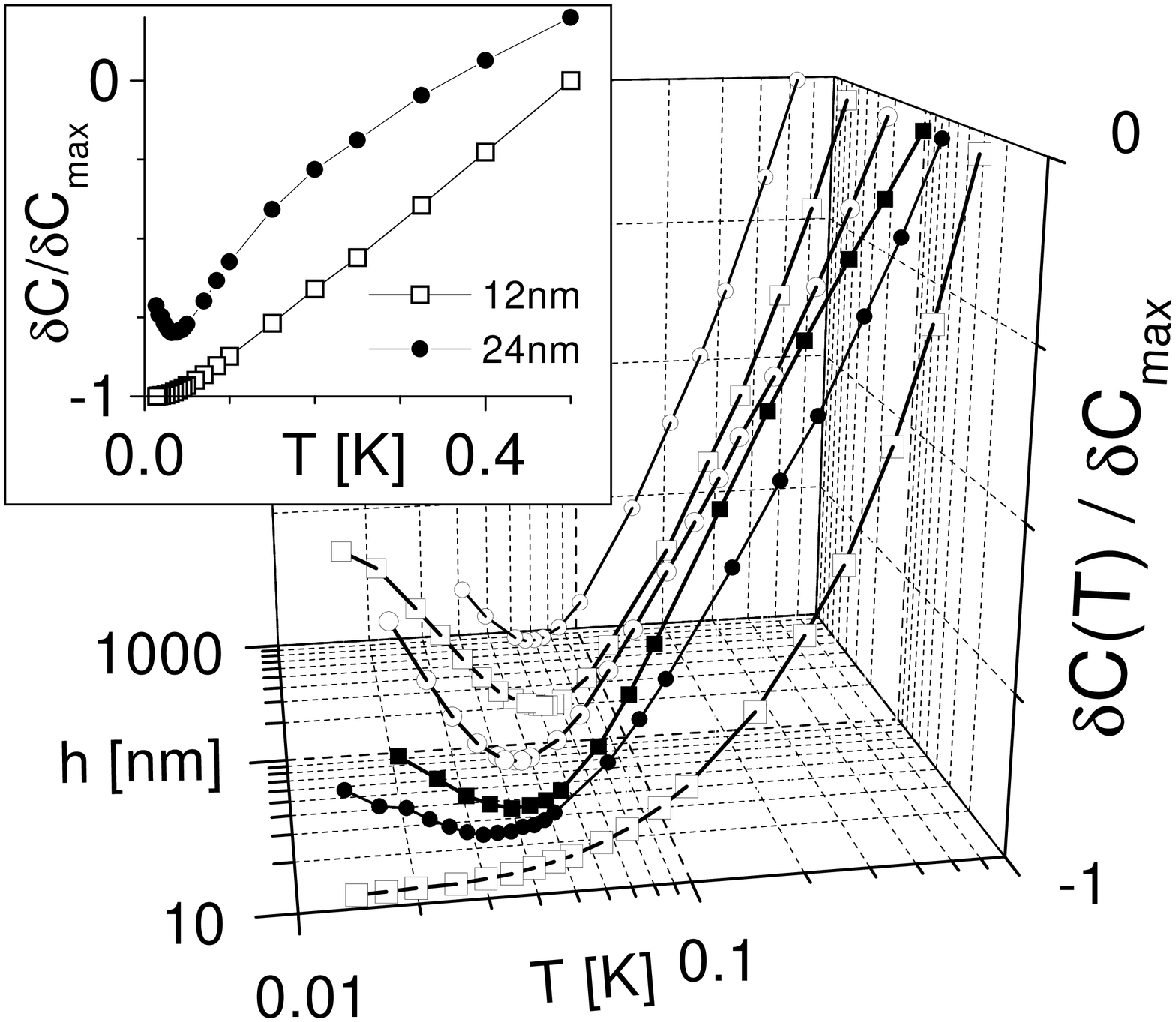}
\caption{ For a given measuring field $E=40$ kV/m, relative variation with $T$  of the 
capacitance: as the sample 
thickness $h$ decreases the minimum shifts towards lower $T$, disappearing for the lowest 
$h$. This contradicts the 
standard \textit{non interacting} two-level 
system model. As the dissipated power ${\cal P} \propto E^2h$ increases with $h$, this $h$ 
effect cannot be due to 
heating by $E$. Inset: Same plot ($T$ in linear scale), the $24$ nm 
curve is vertically moved  ($+0.2$) for clarity. For $h=12$ nm, $\delta C(T)\propto T^{\alpha}$ 
instead of the standard 
$\ln{T}$ behavior obeyed for $h=24$ nm.}
\label{Fig. 1}
\end{figure}

The $\chi'(T,E)$ behavior of six samples, whose $h$ ranges from $12$ nm to $800$ 
nm, are 
reported in Figs. 1-3, where $\delta C (T) = C(T)-C(0.5$ K$)$ and $\delta C_{max} = C(0.5$ 
K$)-C_{min}$ 
while $C_{min}$ is the minimum value of $C$ in our $T$ experimental range $[14$ 
mK$; 0.5$ 
K$]$. Fig. 1 shows that, for a given measuring field $E = 40$ kV/m, the temperature 
$T_{rev}$, where $\chi'(T)$ is 
minimum, gradually moves towards lower values as $h$ is decreased. For the thinnest sample, 
the minimum is suppressed at $E=40$ kV/m, 
and instead of the usual $\delta C(T) \propto \pm \ln{T}$ valid for thicker samples (as well 
as for all previously  
studied glasses \cite{Rogge}), one gets $\delta C(T) \propto T^{\alpha}$ where $\alpha$ turns from $1.35 \pm .15$ 
above $70$ mK to
$2.75 \pm .4$ below. Fig. 2 shows first that this trend of lowering $T_{rev}$ when 
decreasing $h$ holds 
for all $E$ (except for the very few cases $E \ge 1$ MV/m where some heating arises, see 
below); and secondly that this $T_{rev}(h)$ decrease only 
occurs for $h<100$ 
nm since the $200$ nm and $800$ nm samples behave similarly. Last, Fig. 3 shows, for the smallest $h$, that a 
minimum of $\chi'(T)$ is 
recovered 
when $E \gtrsim 1$ 
MV/m, and that the value of the slope below $T_{rev}$ is weakened beside that of thick 
samples.

\begin{figure}
\includegraphics[height=6cm, width=8.5cm]{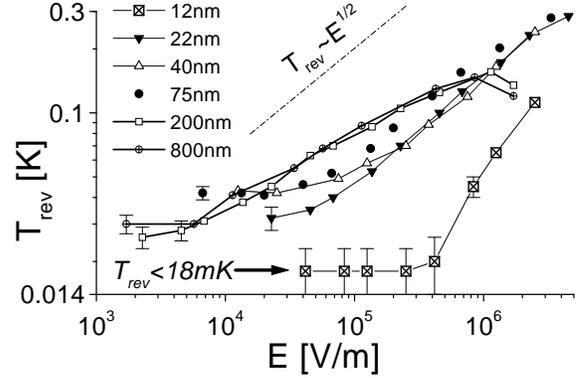}
\caption{ Temperature $T_{rev}$ of the minimum of $\chi'(T)\propto C(T)$, as a function of 
$E$, at various  
$h$ (labelling the curves). For $h=12$ nm, no minimum occurs for $\chi'$ down to $14$ mK, for 
$40$ kV/m$< 
E<250$ kV/m, which is indicated by "$T_{rev}< 18$ mK".  For $E>10$ kV/m, i.e. above the linear 
regime, $T_{rev}$ 
increases with $h$ at any given $E$, except for $E \gtrsim 1$ MV/m where heating occurs for 
the $h>100$ nm 
samples. For clarity, the $\pm 4$ mK uncertainty is shown only for a few lowest $T_{rev}$ .}
\label{Fig. 2}
\end{figure}

These new trends for $\chi'(T,E,h)$ are intrinsic to the glassy state, as we shall see by 
showing that they come 
neither from a heating effect, nor from a variation of the glass composition with $h$:

\textit{i)} the $E$-induced dissipated power $\cal P$ might heat the sample to a temperature 
above the measured $T$. 
Since the thermal conductances decrease as $T$ decreases, heating, at a given $E$, increases at 
low $T$, i.e. it is 
expected to stretch the $\chi'(T)$ curve by an amount increasing as $T$ is lowered. Thus, one 
might wonder whether the 
data of Figs 1-2 are due to heating or not. This explanation is ruled out since, with 
$R\propto h$ the parallel 
resistance of the sample, one has ${\cal P} = E^2 h^2 /R \propto E^2 h$, i.e. $\cal P$ 
\textit{increases} with $h$ at 
a given $E$. Thus, if the data of Figs. 1-2 were mainly due to heating, $T_{rev}$ would decrease for thick samples, 
at odds with Fig. 
1. In fact, heating 
effects are clearly 
visible only for the two thickest samples and for the strongest $E$: $T_{rev}$ is slightly 
lower at $E= 2$ MV/m than at 
$E=1$ MV/m. Assuming that, above the linear regime, i.e. when $E$ is high enough to yield a 
$T_{rev}(E)$ increase, the 
law $T_{rev} \propto E^{1/2}$ is obeyed (see below), one can extract the thermal resistance 
$\cal R$ from the 
difference, increasing with $E$, between this law and the measured $T_{rev}$. From the two 
thickest samples one gets 
${\cal R} \simeq 50$ MK/W at $100$ mK, with a $T^{-3}$ behavior, as expected for boundary Kapitza 
resistances.

\begin{figure}
\includegraphics[height=6cm, width=8.5cm]{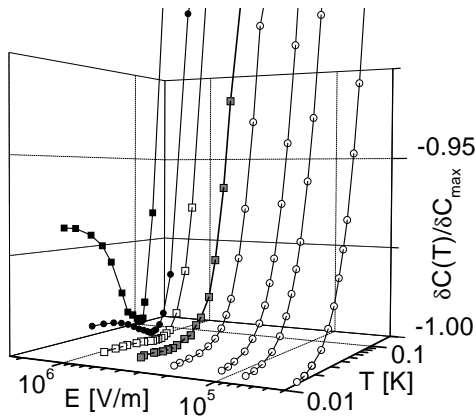}
\caption{ For $h=12$ nm, relative variation of the capacitance with $T$, at various $E$. No 
minimum occurs in 
$\chi'(T)$ down to $14$ mK for the four lowest $E$'s, contrarily to the three highest  
ones. At $E \simeq .4$ 
MV/m (in gray) one may have $T_{rev}=19$ mK. The vertical scale is reduced beside that 
of Fig. 1, since the 
$\chi'(T<T_{rev})$ increase is weaker than in thick samples.}
\label{Fig. 3}
\end{figure}

\textit{ ii)} Besides, one may imagine that the data of Figs. 1-3 come from the fact that the 
glass composition depends on $h$, even for a given set of plasma parameters: due to a possible mixing of the 
plasma incoming particules and of the Cu of the first electrode, one may argue that the samples contain a "boundary" 
layer, of 
thickness $b_0$, whose chemical composition strongly differs of the rest of the glass (of thickness $h-b_0$). Since 
the value of 
$T_{rev}$ 
slightly depends on the 
glass composition \cite{Rogge}, one may wonder whether this could explain the reported data. This is not the case. 
First because 
the $T$ dependence of 
$\delta C(T)$ is 
\textit{totally new} for $h=12$ nm: $\delta C(T) \propto T^{\alpha}$, \textit{clearly 
differs} from the usual 
$\delta C(T) \propto \pm \ln{T}$ behavior holding both for our thicker samples and for all the previously studied 
samples \cite{Rogge} 
whatever their chemical composition. Secondly, 
this scenario leads to add the admittances of the two 
consecutive dielectrics, with the $\chi'$ of the 
boundary layer given 
by the $12$ nm thick sample, and that of the second dielectric by the $h>100$ nm 
samples. Applying this method for the $24$ nm thick sample, at $E=40$ kV/m, one finds first that $T_{rev}$ is 
halved in comparison with thick 
samples, as observed on Fig. 2; and secondly that $\delta C(T)$ should be  
dominated by far by the 
$T^{\alpha}$ behavior, which is \textit{clearly contradicted} by Fig. 1 where the 
$\ln{T}$ trend holds 
for the $24$ nm thick sample.

At this step, it is clear that \textit{decreasing $h$ qualitatively changes the physics of the 
glass sample}. This 
contradicts the standard (i.e. non interacting) TLS model which accounts for the decrease of 
$\chi'$ above $T_{rev}$ 
by the progressive freezing of the diagonal (or relaxationnal) part $\chi'_{z}$ of the 
susceptibility, while the 
$\chi'$ increase below $T_{rev}$ comes from the off-diagonal (or resonnant) part $\chi'_{x}$: 
due to its pure quantum 
nature, $\chi'_x$ grows when $T$ decreases, as do all quantum effects. Within this framework, 
the TLS's interactions 
are assumed to be so small that they only enter in the phase coherence time $\tau_2$ setting 
the scale over which 
$\chi'_{x}$ relaxes towards its thermodynamic value. Since decreasing $h$ reduces TLS's 
interactions (see below), one 
expects qualitatively larger $\tau_2$ in thin samples, reinforcing somehow $\chi'_{x}$, which 
is \textit{at odds with} 
Figs.1-2. Quantitatively, in usual thick samples $\tau_2 \simeq 10\ \mu$s in 
the $20$ mK range, i.e. 
its value is so large that the associated quantum energy $\propto \tau_2^{-1}$ is as low as 
$1\ \mu$K, much smaller 
than any relevant energy scale, which explains that $\chi'_{x}$ does basically not depend on 
$\tau_2$. Thus, 
\textit{no $h$ effect can be explained within the standard TLS model}.

Besides, another serious limit of the standard TLS model, discovered very recently 
\cite{LecochecNL}, is that it does 
not account for the nonlinear $\chi'(T,E)$ usually reported (such as those of our $h>100$ 
nm samples and those of  
Ref. \cite{Rogge}). Indeed, due to the quantum nature of $\chi'$ below $T_{rev}$, $\chi'$ is 
strongly 
\textit{depressed} by a strong measuring electric field $E$ at a given $T$. This is due to the 
\textit{quantum 
saturation} phenomenon coming from the fact that increasing $E$ decreases the population 
difference between the 
two energy levels: as the Rabi oscillations produced by $E$ on the upper level are in phase 
opposition with respect to 
those produced on the ground level, the quantum response, once averaged on many independent 
TLS's, tends to zero 
when $E$ is increased. This was checked by solving numerically the Bloch equations of TLS's 
\cite{LecochecNL} with a 
non perturbative method (see the dotted lines in Fig. 4).

\textit{Both} the nonlinear measurements as well as the $h$ effect on $\chi'$ can be 
accounted for by using the 
\textit{same} $E$-induced TLS-TLS interaction proposed by Burin \textit{et al.} \cite{Burin} where thermal 
excitations, which 
are at zero-field 
localized on each TLS, tend to delocalize by hopping to resonant nearest neighbors. This is 
due to the fact that 
\textit{resonant hopping} demands that \textit{both} TLS's should have approximately the same 
asymmetry energy 
$\Delta$ \textit{and} the same tunneling energy 
$\Delta_{0}$: as the electrical field modulates the TLS parameter $\Delta$, the probability of 
finding, for a given 
TLS, a resonant TLS,  increases from a negligible value at very low $E$, to a non-negligible 
value above at higher $E$. Since this mechanism transports energy, it mainly enhances the diagonal part $\chi'_z$ 
of the susceptibility 
(by decreasing the associated relaxation time $\tau_{1}$), and since it gets stronger as $T$ 
decreases it yields the 
trend shown in Fig. 4 by the solid curves: in this picture, $\chi'$ mainly 
comes from $\chi'_z$ which 
dominates, \textit{at any} $T$, over the off diagonal part $\chi'_x$ (the latter being still depressed by the $E$ 
values used in 
experiments). With $\tau_{1} \propto T/\sqrt{E}$, see Ref. \cite{LecochecNL}, \cite{Neu}, 
Burin's mechanism is 
essential at low $T$, and is responsible for the $\chi'(T)$ increase below $T_{rev}$ when $\omega 
\tau_1$ becomes smaller 
than unity, with $\omega$ the frequency of $E$. This yields $T_{rev} \propto \sqrt{E}$, 
\textit{as seen on Fig. 2}.

\begin{figure}
\includegraphics[height=6.7cm, width=9.1cm]{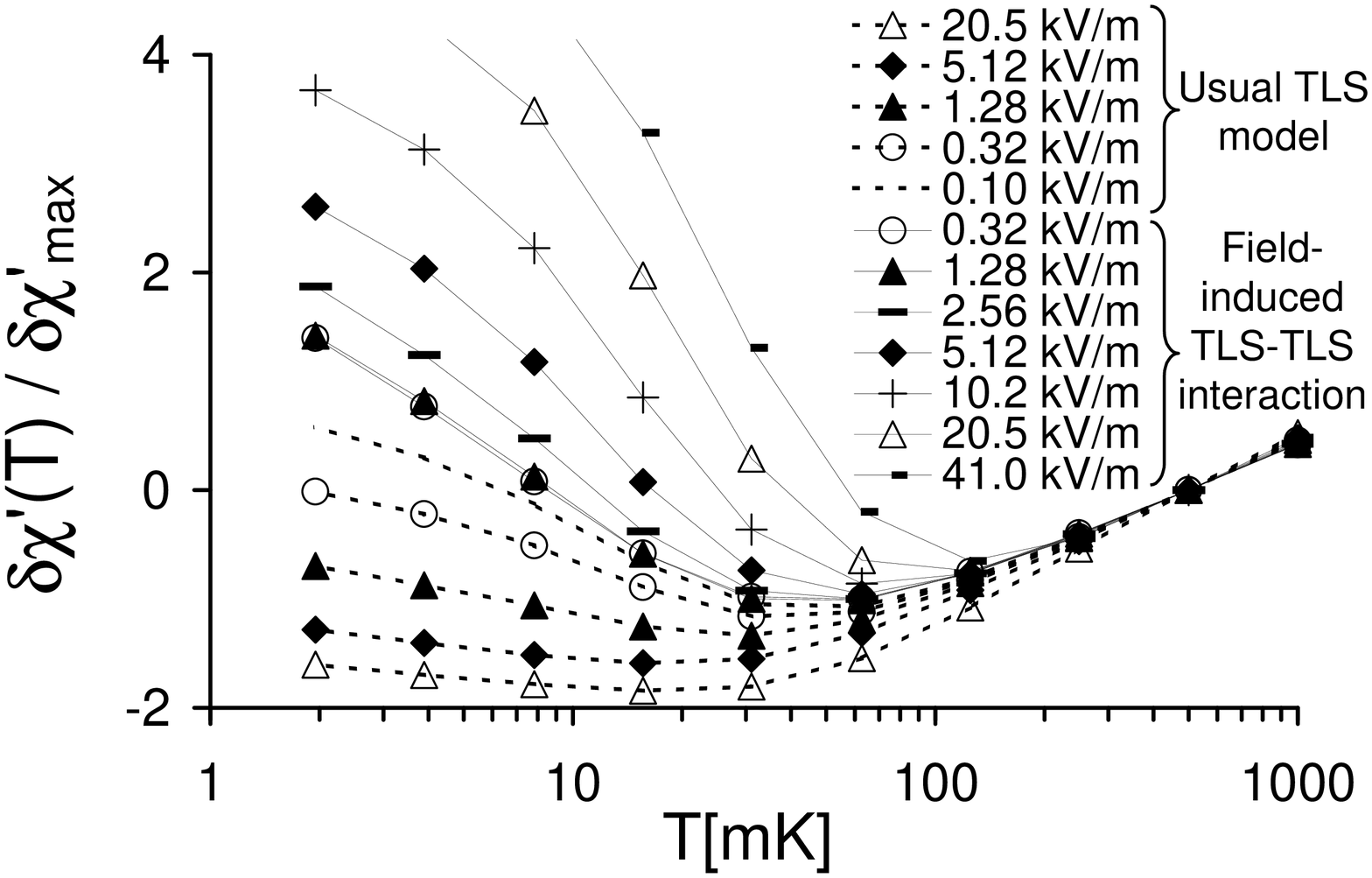}
\caption{Solving Bloch equations, $\chi'(T,E)$ was simulated in Ref. \cite{LecochecNL}.  The 
non interacting TLS model 
yields the dotted lines: due to the quantum saturation effect, $\chi'\propto C$ is strongly 
depressed when rising $E$ 
at  
low $T$, suppressing the minimum of $\chi'(T)$ for experimental fields $E > 40$ kV/m and 
modelling the $12$ nm 
sample where TLS's interactions are strongly reduced. Including TLS's interactions through the 
$E$-induced 
mechanism of Ref. \cite{Burin} restores the $\chi'$ increase with $E$ at any given $T$, as well as that of 
$T_{rev}$: this models the usual $\chi'(T,E)$ of thick samples. Since strong $E$'s decrease the 
distance $\lambda$ 
between 
interacting TLS's, one recovers the usual behavior of $\chi'(T)$ when 
$\lambda (E)<h$: this explains, for $h=12$ nm, the recovery of a $\chi'(T)$ minimum for strong 
$E$, as in Figs 
2-3.}
\label{Fig. 4}
\end{figure}

In this picture, $h$ effects on $\chi'$ are due to the two mechanisms reducing TLS's interactions at small $h$'s. 
First, due to the 
usual density of states \cite{Philipps}, if $w$ is the maximum energy separating the gaps of two interacting TLS's,  
they are 
separated by the distance $\lambda \propto w^{-1/3}$ \textit{only while} $\lambda <h$. For thin samples $h<\lambda$, 
the decrease of 
available TLS's yields $\lambda \propto w^{-1/2}$ and \textit{enlarges} $\lambda$, reducing TLS's interactions $U 
\propto 
\lambda^{-3}$. This applies both to the elastic part of $U$ where \cite{Neu} one has $w \simeq T$, and to the 
dipolar part of $U$ 
where Burin's mechanism states $w=e_{dip}= pE/\epsilon_r$ with $p\simeq 1$ D the TLS dipole and $\epsilon_r \simeq 
5$ the glass 
dielectric constant \cite{Carruzzo}, \cite{Rogge}. Secondly, dipolar interactions are further \cite{note} reduced by 
the screening 
effect in the electrodes: if $h<\lambda$ their numerous electrons intercept and cancel the electric field yielding 
the interaction 
between TLS's, which sharply decreases $U$. 

Finally, for $E=40kV/m$ one gets $e_{dip}=10$ mK, not far from the $T$ range where the $\chi'(h)$ effects occur in 
Figs 1-2. Whatever 
$w=T$ or $w=e_{dip}$, the order of magnitude of $\lambda (w)$ is near $60$ nm for thick samples: \textit{this is 
consistent with the 
thickness where the $\chi'(h)$ effects occur in Figs. 1-2}. Since TLS's interactions bring about the 
$\chi'(T<T_{rev})$ behavior, 
their suppression at small $h$'s should yield a $\chi'(T)$ curve given by the standard \textit{non interacting} TLS 
model: as shown by 
the dotted lines in Fig. 4, this amounts to suppressing the $\chi'(T)$ minimum, \textit{as in Fig. 1} for $h=12$ nm, 
due to the 
$E$-induced strong depression of 
$\chi'_{x}$ for the $E>40$ kV/m experimental fields. 

Besides, due to the 
$E$-induced modulation of $\Delta$, in Burin \textit{et al}.'s scenario $\lambda$ decreases at strong 
$E$, opening the 
possibility to recover a $\lambda <h$ case at strong $E$: \textit{this would account for Fig. 
3} where, even for 
$h=12$ nm, one recovers a minimum for $\chi'(T)$. 
  
In conclusion, decreasing the thickness of glass samples down to the $10$ nm range changes the 
physics of the real 
part of the kHz dielectric susceptibility. This strongly evidences the key role of TLS's 
interactions up to $100$ 
mK. Assuming that TLS's interactions occur via a $E$-induced delocalisation of 
excitations between quasi 
similar TLS's accounts for the main features of the reported data.

Many thanks to M. Ocio and E. Vincent for lending us their cryostat and to L. Le Pape for 
experimental support. A special thank to J.-Y. Prieur and 
J. Joffrin (CNRS, Orsay) for the really encouraging scientific discussions we had, 
and to D.L'H\^ote for experimental 
help.


%
%


\begin{thebibliography}{99}
\bibitem{note} If screening is the major reason of the decrease of $U$, one could explain why no $h$ effect was 
seen, in the nm range, 
 in an internal kHz friction experiment \cite{White}: electrodes do not exist in that case.
\end{thebibliography}
\end{document}